\documentclass[%
 reprint,		
 amsmath,amssymb,superscriptaddress,
 aps,
 prmaterials,
]{revtex4-2}

\usepackage{graphicx}
\usepackage{dcolumn}
\usepackage{bm}
\usepackage{float}			
\usepackage{hyperref}
\hypersetup{colorlinks=true, citecolor=blue, urlcolor=blue, linkcolor=blue}
\begin{document}

\preprint{APS/123-QED}

\title{Superionic State Discovered in Ternary Hypervalent Silicon Hydrides via Sodium inside the Earth}
\author{Tianxiao Liang}	
\affiliation{College of Physics, Jilin University, Changchun 130012, China}
\author{Zihan Zhang}	
\affiliation{College of Physics, Jilin University, Changchun 130012, China}
\author{Hongyu Yu}	
\affiliation{College of Physics, Jilin University, Changchun 130012, China}
\author{Tian Cui}	
\affiliation{College of Physics, Jilin University, Changchun 130012, China}
\affiliation{Institute of High Pressure Physics, School of Physical Science and Technology, Ningbo University, Ningbo 315211, China}
\author{Chris J. Pickard}	
\affiliation{Department of Materials Science \& Metallurgy, University of Cambridge, 27 Charles Babbage Road, Cambridge CB3 0FS, United Kingdom}
\affiliation{Advanced Institute for Materials Research, Tohoku University 2-1-1 Katahira, Aoba, Sendai, 980-8577, Japan}
\author{Defang Duan}	
\email{duandf@jlu.edu.cn}
\affiliation{College of Physics, Jilin University, Changchun 130012, China}

\begin{abstract}
Superionic states are phases of matters that can simultaneously exhibit some of the properties of a fluid and of a solid. Superionic states of ice, H$_{3}$O, He-H$_{2}$O or He-NH$_{3}$ compounds have been reported in previous works. Silicon, sodium, and hydrogen are abundant elements inside the earth. Here, we use ab initio calculations to show that, at extreme conditions inside the earth, Na, Si, and H can form many hypervalent compounds that some of them can exist every close to ambient pressure, and surprisingly a previously unknown type of superionic state of $P\overline3m1 - $Na$_{2}$SiH$_{6}$ can form as well. Our work focused on new superionic state of Na$_{2}$SiH$_{6}$, and the results also reveal several different hypervalent Si-H anions discovered, which are different from individual SiH$_{5}^{\,-}$ and octahedral SiH$_{6}^{\,2-}$ in previous research of ternary alkali hypervalent silicon hydrides. Our work provides some advice on further investigations on potential ternary hydrides inside the earth.
\end{abstract}

\maketitle
\section{Introduction}
The pioneering work of considerable effort has been devoted to studying superionic states in “hot ice” layers of Uranus or Neptune \cite{ryzhkin1985superionic,demontis1988new,cavazzoni1999superionic,schwegler2008melting} and several helium compounds at high pressure reported by Liu \emph{et al} \cite{liu2019multiple,liu2020plastic}. Huang \emph{et al} also proposes the possible existence superionic H$_{3}$O in Uranus or Neptune by structural prediction \cite{huang2020stability}. It is believed that planetary environment are conducive to the formation of superionic states \cite{french2009equation,redmer2011phase,wilson2013superionic,sun2015phase}. Experimental electrical conductivity measurements \cite{yakushev2000electrical,chau2001electrical} and Raman signals \cite{goncharov2005dynamic} of water under compression indicate the existence of superionic states. Silicon(Si) is the second most abundant element, which is an important component of the earth crust. And sodium(Na) is the seventh most abundant element in the earth crust. There are high abundance, which are estimated about $28.2\%$ for Si and $2.36\%$ for Na in earth crust from CRC. Besides that hydrogen has an abundance of $0.14\%$ by weight and the third most abundant by number of atoms. At room temperature and ambient pressure, pure hydrogen exists as a diatomic gas (H$_{2}$). There is a small amount of hydrogen gas in the earth’s atmosphere which makes up less than one part per million and is abundant in compound form inside the earth. At deep-earth conditions (high pressure and high temperature), our research suggests that Na-Si-H hypervalent compounds will form inside the earth. 

Hypervalences are well established in chemistry, which central atoms attain an environment of valence electrons that exceeds the number of eight through their coordination with ligands \cite{treichel1967synthesis,noury2002chemical}. Bonding in hypervalent species in straight line is visually described by the three-center-four-electron (3c-4e) model \cite{doi:10.1021/ja01153a086,pimentel1951bonding,koutecky1974molecular}. And the Lewis-Langmuir theory of valence attributed the stability of molecules to their ability to place their valence electrons, which appropriately paired off as chemical bonds into stable octets \cite{lewis1916atom,langmuir1919arrangement}. Some previous theoretical and experimental researches show that alkali metal potassium(K) could form K-Si-H compounds, in which potassium silyl K$_{2}$SiH$_{6}$ with octahedral-structured SiH$_{6}^{\,2-}$ anions maintain excellently reversible reactions through direct hydrogenation of the KSi Zintl phase at ambient conditions, consistent with theoretical calculated results \cite{janot2016catalyzed,chotard2011potassium,ring1961preparation,ring1961crystal}. Accordingly, we believed other that alkali metal elements and hypervalent Si-H anions can form different ternary compounds in the formation of A$_{2}$SiH$_{6}$, (A indicates Li or Na), or other formations. In recent work, Liang \emph{et al} \cite{liang2020ternary} and Zhang \emph{et al} \cite{zhang2020structure} come up with several different types of Li-Si-H compounds, respectively, including layer typed SiH$_{5}^{\,-}$ in $R32$-LiSiH$_{5}$, tricapped triangular prismatic SiH$_{6}^{\,2-}$ in $P\overline62m$-Li$_{2}$SiH$_{6}$, and hydrogen rich compounds Li$_{3}$SiH$_{10}$ and Li$_{2}$SiH$_{6+\delta}$ by Liang \emph{et al}. And hydrogen poor Li-Si-H superconductors by Zhang \emph{et al}.

In our work, many ternary Na-Si-H phases were explored with peculiar structures and properties at high pressure, and we try to realize more about the formation of hypervalent Na-Si-H compounds. In addition, a new superionic phase of $P\overline3m1 - $Na$_{2}$SiH$_{6}$ was discovered at extreme environment inside the earth. We come up with the probable phase diagram of $P\overline3m1 - $Na$_{2}$SiH$_{6}$ and potential synthesis areas inside the earth. And it is possible to observe the superionicty by experimental methods.

	\begin{figure*}[htbp]
	\centering
	\includegraphics[width=\textwidth]{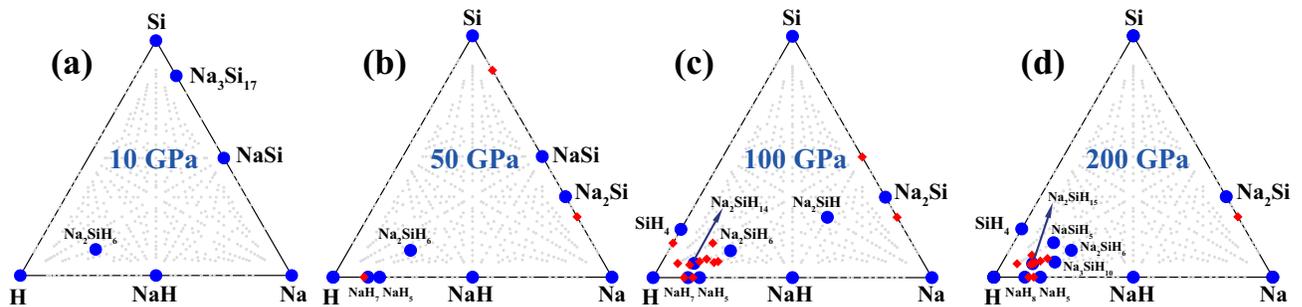}
	\caption{The ternary phase diagram of Na-Si-H at (a) 10 GPa, (b) 50 GPa, (c) 100 GPa, and (d) 200 GPa. Predicted structural parameters of ternary compounds are shown in Table S1 in supplementary material. Big purple circles represent compounds which located on the convex hulls, and small red diamonds represent which didn't locate on the convex hulls.}
	\label{figure:CH}
	\end{figure*}
	
\section{Computational Details}
Structure searches of Na-Si-H was performed within \emph{ab initio} calculations as implemented in the AIRSS (Ab Initio Random Structure Searching) codes \cite{pickard2011ab,pickard2006high} combining with CASTEP code \cite{segall2002first}. Full variable-composition predictions were firstly performed within 50,000 structures at pressures of 10 GPa, 50 GPa, 100 GPa, and 200 GPa. Then, fix composition predictions were employed to further search structures for stable compounds. The VASP (Vienna ab initio simulation packages) code \cite{kresse1996efficiency} was used to optimize crystal structures and calculate the electronic properties, where the Perdew–Burke–Ernzerhof (PBE) \cite{perdew1996generalized} of generalized gradient approximation (GGA) \cite{perdew1992pair} with the all-electron projector-augmented wave method (PAW) \cite{blochl1994projector} was performed. The electronion interaction was described by projector-augmented-wave potentials with the $1s^{1}$, $2s^{2}2p^{6}3s^{1}$ and $2s^{2}2p^{6}3s^{2}3p^{2}$ configurations treated as valence electrons for H, Na and Si, respectively. The Perdew-Burke-Ernzerhof generalized gradient approximation was chosen for the exchange-correlation functional, and kinetic cutoff energy of $800$ eV, and Monkhorst-Pack $\bm{k}$ meshes with grid spacing of $2\pi \times 0.03 \mathring{A}^{-1}$ were then adopted to ensure the enthalpy converges to less than $1$ meV/atom. Ground-state and some of semi-state structures for Na-Si-H compounds at 50 GPa, 100 GPa, and 200 GPa were chosen in the calculations of convex hull. The phonon calculations were as implemental in the PHONOPY code \cite{togo2008first}, which the computational settings described below. The Bader charge analysis \cite{bader1975acc} was used to determine charge transfer, and the electron localization function (ELF) \cite{becke1990simple} was used to describe and visualize chemical bonds in molecules and solids. 

For more dynamical properties, we have performed extensive \emph{ab initio} molecular dynamics (AIMD) simulations based on the Born-Oppenheimer approximation implemented in VASP \cite{xia2018novel} within the pressure range $3 \sim 200$ GPa and set the temperature range from 1,000 K to 3,000 K,  And we extended several points to 30 ps to check the stability of the simulations. $P\overline3m1 - $Na$_{2}$SiH$_{6}$ was performed in $3 \times 3 \times 3$ supercells (with 243 atoms). Both $1\times1\times1$ and $2\times2\times2$ $\bm{k}$-mesh are selected. $R\overline3m$-NaSiH$_{5}$ was performed in $3 \times 3 \times 1$ supercells (with 189 atoms), and others were performed in $2 \times 2 \times 3$ or $2 \times 3 \times 3$ supercells. We adopted the canonical $NVT$ ensemble, lasting for 12 ps with a time step of 1 fs, and we allowed 2 ps for thermalization and then extracted data from the last 10 ps. Some simulations were extended to 20 ps to check the stability. MDAnalysis codes \cite{gowers2019mdanalysis,michaud2011mdanalysis} are used to postprocess the AIMD results.

\section{Results and Discussions}
	\begin{figure}[htbp]
	\centering
	\includegraphics[width=0.5\textwidth]{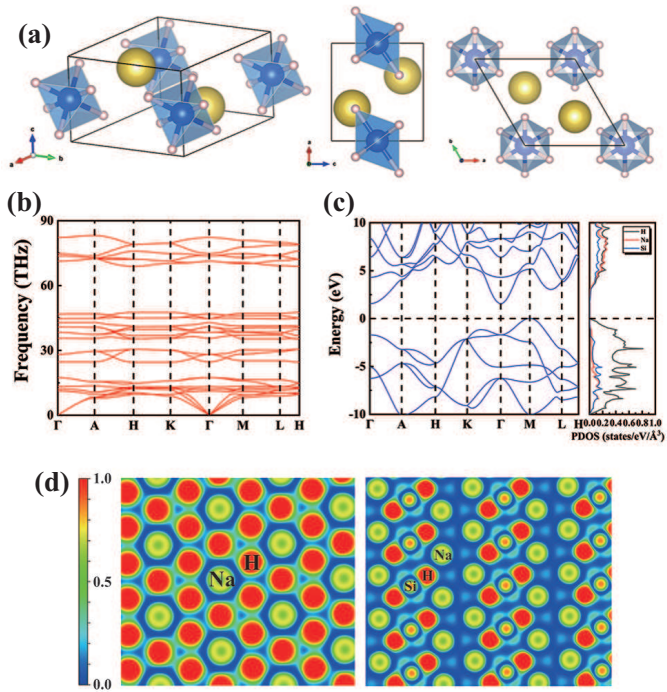}
	\caption{Calculations of properties of $P\overline3m1 - $Na$_{2}\text{SiH}_{6}$. (a) Predicted $P\overline3m1 - $Na$_{2}\text{SiH}_{6}$ structure. (b) Phonon dispersion relations at 100 GPa. The absence of imaginary frequency in the phonon spectra indicates the structures are dynamically stable. (c) Electronic band structures and projected density of states at 100 GPa. More results of $P\overline3m1 - $Na$_{2}\text{SiH}_{6}$ are listed in supplementary material. (d) ELF calculation in (0, 0, 1) and (2, -1, 0) sections.}
	\label{figure:Na2SiH6}
	\end{figure}

We have performed an extensive exploration of the high pressure phase diagrams of Na-Si-H system using random structure searching method implemented in the AIRSS code. This has led us to predict five stable stoichiometries, Na$_{2}$SiH, NaSiH$_{5}$, Na$_{2}\text{SiH}_{6}$, Na$_{3}\text{SiH}_{10}$, and Na$_{2}\text{SiH}_{14}$, as shown in Fig. \ref{figure:CH}(a). At 10 GPa, only Na$_{2}\text{SiH}_{6}$ is stable located on the convex hull. As pressure increasing, more and more stoichiometries can be stable and there are NaSiH$_{5}$, Na$_{2}\text{SiH}_{6}$, Na$_{3}\text{SiH}_{10}$, and Na$_{2}\text{SiH}_{15}$ located on the convex hull at 200 GPa. More interesting, Na$_{2}\text{SiH}_{6}$ is the only one can be stable in the pressure range of 10-200 GPa. It should be noted that there are several energetically competitive structures for these type of molecular crystals.

Na$_{2}\text{SiH}_{6}$ adopts hexagonal structure of $P\overline3m1$ symmetry, in which Si bond with six H forming octahedral SiH$_{6}^{\,2-}$ listed in Fig. \ref{figure:Na2SiH6}(a). And there is no phase transition as pressure up to 200 GPa. As mentioned in Fig. S1 in the supplementary material (SM) \cite{SupplementalMaterial}, $P\overline1$-Na$_{3}\text{SiH}_{10}$ and $P\overline1$-Na$_{2}$SiH$_{6+\delta}$ are all formed based on octahedral SiH$_{6}^{\,2-}$ anions. For Na$_{2}\text{SiH}_{2+\delta}$ and Na$_{3}\text{SiH}_{10}$, there are many H$_{2}$ units extended into different directions in their cells reducing their symmetries to $P\overline1$, and all of hydrogen-rich compounds are hypervalences. Structural parameters are listed in Table S1 in the SM \cite{SupplementalMaterial}. And results for other properties are shown in Fig. S3 $\sim$ Fig. S23 in the SM \cite{SupplementalMaterial}. And $N-X-L$ notations \cite{perkins1980electrically} for Na-Si-H compounds are listed in Table S2 in the SM \cite{SupplementalMaterial}.

We examine the electronic properties of Na-Si-H compounds, and Bader charge analysis reveals a charge transfer from Na/Si to H listed in Table S3 in the SM \cite{SupplementalMaterial}, suggesting that both Na and Si are electron donors and provide electrons to H atoms. Na constantly loses approximate $0.75 \sim 0.81$e per atom and Si loses approximate $2.6 \sim 3.0$e per atom in stable phases. And Na-ions bond covalent bonds with Si-H anions. Bonding is covalent between H and Si atoms to compose hypervalent silicic anions. Almost all of Na-Si-H compounds are semiconductors or insulators with energy gaps higher than $1.0$ eV. It is shown that all the predicted compounds have high gravimetric hydrogen contents and volumetric hydrogen densities, suggesting that they may be potential hydrogen storage materials. 

	\begin{figure*}[htbp]
	\centering
	\includegraphics[width=0.9\textwidth]{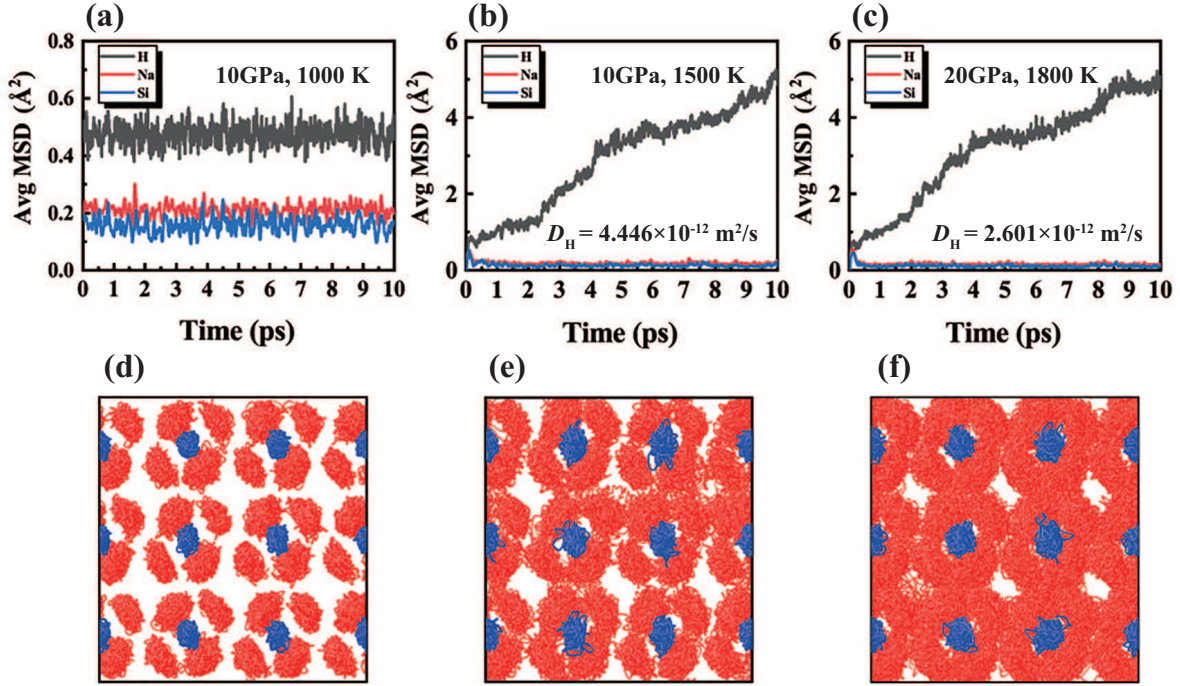}
	\caption{Dynamical behaviors of H (green) and Na atoms (orange) compared to Si atoms (blue) in the Na-Si-H compounds under high pressure and temperature. MSD calculated from the AIMD simulations for $P\overline3m1 - $Na$_{2}\text{SiH}_{6}$ compounds in the solid and superionic states. (a) $\sim$ (c) are the averaged mean-squared displacements (MSD) under different pressures and temperatures. The diffusion was inhibited rather than enhanced by high temperature and pressure. (d) $\sim$ (f) are the representations of atomic trajectories in one supercell from the simulations from the last 10 ps run representing the three distinct phases: (d) the solid phase (10 GPa and 1,000 K), (e) the superionic phase (10 GPa and 1,500 K), (f) the superionic phase (20 GPa and 1,800 K). Only (1, 0, 0) Si-H layers are plotted in (d) $\sim$ (f), and the Si and H atoms are dividedly plotted in blue and red.}
	\label{figure:AIMD}
	\end{figure*}
	
	\begin{figure*}[htbp]
	\centering
	\includegraphics[width=0.9\textwidth]{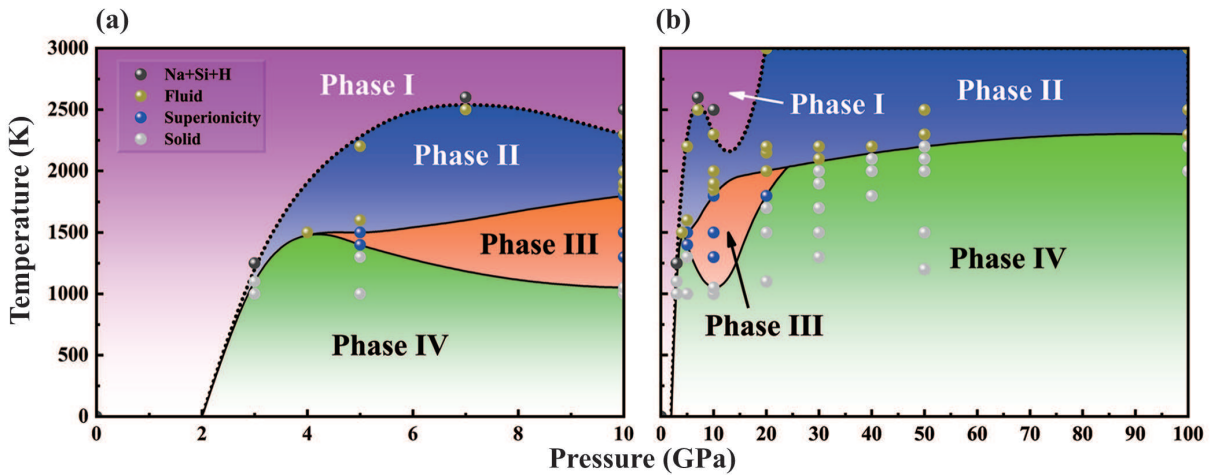}
	\caption{Proposed phase diagram of $P\overline3m1 - $Na$_{2}$SiH$_{6}$ at high pressures obtained from structure searches and AIMD simulations. Different colored balls are used to mark the simulations. Black solid and dash lines are fitted to the phase transition boundaries plotted by Gibbs free energy. Phase I indicates decompositional state with Na, Si, and H. Phase II indicates the fluid of Na$_{2}$SiH$_{6}$. Phase III indicates the superionic state of Na$_{2}$SiH$_{6}$. And Phase four indicates the solid of Na$_{2}$SiH$_{6}$.}
	\label{figure:PT}
	\end{figure*}

	\begin{figure}[htbp]
	\centering
	\includegraphics[width=0.5\textwidth]{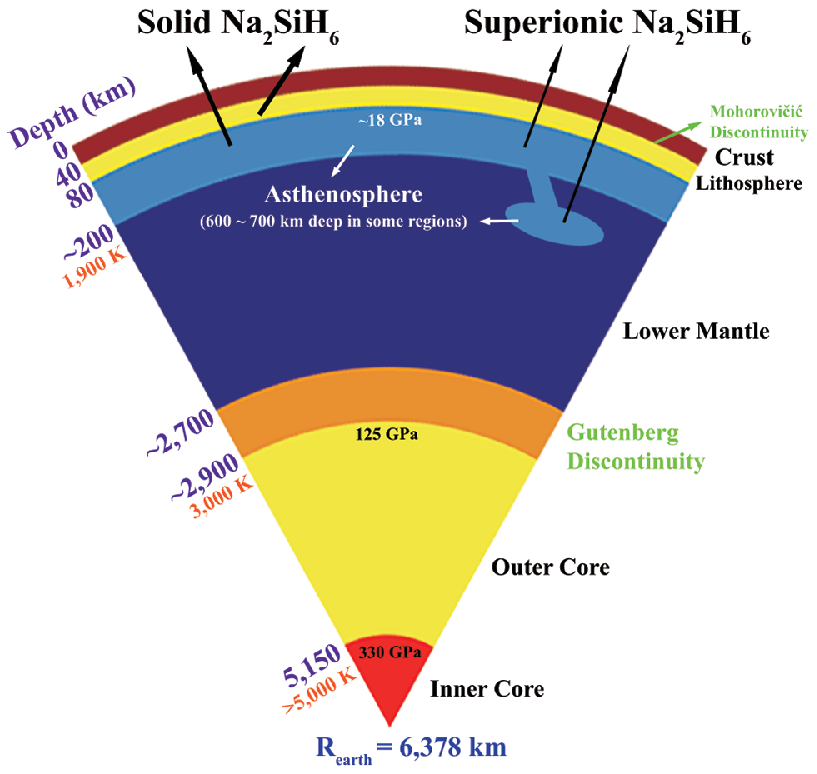}
	\caption{Layered structure of earth's interior along with depth, pressure, and temperature, and potential synthesis areas of $P\overline3m1 - $Na$_{2}$SiH$_{6}$. Crust and mantle are divided by Mohorovičić discontinuity, and Na$_{2}$SiH$_{6}$ may exist in lithosphere or asthenosphere. The depth of different layers are not exactly the measuring scale, just for ease of observation.}
	\label{figure:earth}
	\end{figure}
				
Next, we study more dynamical properties of the predicted Na-Si-H compounds and investigate the diffusion of atoms, as shown in Fig. \ref{figure:AIMD}. Atomic diffusion is plotted by mean-square displacements (MSDs) at the temperature range 1,000 $\sim$ 3,000 K. At 1,000 K, the vibration of atoms for NaSiH$_{5}$, Na$_{3}$SiH$_{10}$, and Na$_{2}$SiH$_{14}$ located on the convex hull are weak and atoms cannot jump to other neighbouring sites. All the atoms vibrate nearby the equilibrium positions, see Fig. S24 in the SM \cite{SupplementalMaterial}. And SiH$_{5}^{\,-}$ or SiH$_{6}^{\,2-}$ ions will not decompose at this temperature. As mentioned above, NaSiH$_{5}$, Na$_{3}$SiH$_{10}$, and Na$_{2}$SiH$_{14}$ can be classified as solid at 1,000 K. Surprisingly, in our Na-Si-H compounds, we found a region of superionicity with an extraordinary diffusive H phase of $P\overline3m1 - $Na$_{2}$SiH$_{6}$. Superionic $P\overline3m1 - $Na$_{2}$SiH$_{6}$ can be classified as one of three phases in terms of the diffusion coefficient ($D$) of the H atoms:  the solid phase ($D_{\text{H}} = D_{\text{Na}} = D_{\text{Si}} = 0$), the fluid phase ($D_{\text{H}} > 0$, $D_{\text{Na}} > 0$, and $D_{\text{Si}} > 0$), and the superionic (SI) phase ($D_{\text{H}} > 0$ but $D_{\text{Na}} = D_{\text{Si}} = 0$). $D$ were calculated for the hydrogen atoms from their MSDs values. Our simulations show that H atoms have different ‘melting’ temperatures different from Na or Si atoms, and it tended to be superionicity at the temperature range of 1,050 $\sim$ 1,800 K. At about 5 GPa and 1400 K,  the superionicity of $P\overline3m1 - $Na$_{2}$SiH$_{6}$ was firstly observed. At 10 GPa, $P\overline3m1 - $Na$_{2}$SiH$_{6}$ tended to be solid ($D_{\text{H}} = 0$) till the temperature up to 1,050 K. And with the temperature increasing above 1,800 K, Na$_{2}$SiH$_{6}$ tended to be fluid. As listed in Table S4 in the SM \cite{SupplementalMaterial}, high pressure inhibited the diffusion of H atoms but high temperature enhance the diffusion of H atoms, which is specific in the values of $D_{\text{H}}$ in  Fig. \ref{figure:AIMD}. Inspired by the temperature-induced features in $P\overline3m1 - $Na$_{2}$SiH$_{6}$ discussed above, we expanded the pressure range studied to explore the superionic region from ambient pressure up to about 100 GPa with pressure steps of 10 GPa, and calculated from ambient pressure up to 10 GPa in details with pressure steps of 1 GPa. Each of colored balls corresponds to an independent simulation. As shown in Fig. \ref{figure:PT}, two different melting curves divide its phase diagram into three regions: solid, superionicity and fluid. The thermal fluctuations of $P\overline3m1 - $Na$_{2}$SiH$_{6}$ increase rapidly at low pressure range. In Phase III (the superionic state), the thermal fluctuations increase slowly above 20 GPa. Finally, the superionicity is inhibited with increasing pressure and finally disappears at higher pressure. And the melting curve divided phase II and phase IV tend to be flat above 30 GPa. The temperature may therefore have a critical influence on the transition pressure Phase II, Phase III, and Phase IV phases at high pressure. Law of equipartition indicates that the energy should be equally distributed in equilibrium among the degrees of freedom. Thus, the lighter atoms with the same kinetic energy should move faster in non-interacting systems. Within our Na-Si-H compounds, H atoms from SiH$^{2-}_{6}$ ions get a greater mobility. Combining with Bader analysis results in Table S3, radial distribution functions (RDF) calculated in Fig S25, and motion behaviors of atoms of $P\overline3m1 - $Na$_{2}$SiH$_{6}$ in Table S4 in the SM \cite{SupplementalMaterial}, it should be H$^{-}$ of SiH$^{2-}_{6}$ ions that move faster, and H$^{-}$ ions should play an important role in charge transport, which is different from H$^{+}$ in the pioneering work. Temperature and pressure required for synthesis of superionic state of $P\overline3m1 - $Na$_{2}$SiH$_{6}$ are much better than superionic water, H$_{3}$O, or He-H$_{2}$O/He-NH$_{3}$ compounds \cite{ryzhkin1985superionic,demontis1988new,cavazzoni1999superionic,schwegler2008melting,liu2019multiple,huang2020stability,liu2020plastic}, which it may be obsered by measuring the electrical conductivity of H$^{-}$ in SiH$^{2-}_{6}$ \cite{yakushev2000electrical,chau2001electrical}. Recommended synthetic routes of $P\overline3m1 - $Na$_{2}$SiH$_{6}$ are sodium silicide (Na$_{x}$Si$_{y}$) reacted with hydrogen, or sodium hydride (NaH) reacted with silicon and hydrogen under high pressure and temperature.

Comparing with the environment inside the earth, as shown in Fig. \ref{figure:earth}, solid or superionic $P\overline3m1 - $Na$_{2}$SiH$_{6}$ may exist in the upper mantle. Solid $P\overline3m1 - $Na$_{2}$SiH$_{6}$ may exist in lithospheric mantle (lithosphere), which is solid and ridged shell under crust. Asthenosphere, as a weak, uneven, and mushy shell, is another part of the upper mantle. The upper asthenosphere close to lithosphere with lower temperature may contain solid $P\overline3m1 - $Na$_{2}$SiH$_{6}$ as well. And superionic $P\overline3m1 - $Na$_{2}$SiH$_{6}$ may exist in lower asthenosphere.

\section{Conclusion}
In summary, our crystal structure searches combined with property calculations have identified several hypervalent structures, which are dynamically stable at high pressure condition to simulate the environment inside the earth by employing the within \emph{ab initio} calculations as implemented in the AIRSS within first-principles calculations, like $R\overline3m - $NaSiH$_{5}$, $P\overline1 - $Na$_{3}$SiH$_{10}$, $P\overline3m1 - $Na$_{2}$SiH$_{6}$, $Imm2 - $Na$_{2}$SiH$_{12}$, and $P\overline1 - $Na$_{2}$SiH$_{6+\delta}$. Several different types of hypervalent ions discovered in Na-Si-H ternary compounds, which are layer-typed SiH$_{5}^{\,-}$ and linear SiH$_{6}^{\,2-}$. For $P\overline3m1 - $Na$_{2}$SiH$_{6}$, we find the superionicity at high pressure and temperature. At relatively low pressures (below 5 GPa), the superionic region exists and is inhibited by high temperature above 10 GPa, and finally disappear. We have come up with two potential synthetic routes of $P\overline3m1 - $Na$_{2}$SiH$_{6}$. The pressure and temperature ranges are more proper comparing with hot ice or other superionic compounds, which may be more easily accessed in future experiments. The solid and superionic state of $P\overline3m1 - $Na$_{2}$SiH$_{6}$ may occur inside the upper mantle of the earth. All of the above indicate that abundant types of Na-Si-H ternary phases could form inside the earth at high pressure and high temperature. We believe that our work provides a new perspective on deep geological structures of the earth or other terrestrial planets, and finding more superionic states or other interesting properties.

\begin{acknowledgments}
This work was supported by National Natural Science Foundation of China (Nos. 11674122 and 11704143). CJP acknowledges financial support from the Engineering and Physical Sciences Research Council [Grant EP/P022596/1] and a Royal Society Wolfson Research Merit award. Parts of calculations were performed in the High Performance Computing Center (HPCC) of Jilin University and TianHe-$1$(A) at the National Supercomputer Center in Tianjin.
\end{acknowledgments}


\end{document}